\documentstyle[12pt,epsfig,amsmath,amssymb]{article}
\begin{document}

\title{\bf Classical system boundaries cannot be determined within quantum Darwinism}

\author{{Chris Fields}\\ \\
{\it 21 Rue des Lavandi\`eres}\\
{\it Caunes Minervois 11160 France}\\ \\
{fieldsres@gmail.com}}
\maketitle

\begin{abstract}
Multiple observers who interact with environmental encodings of the states of a macroscopic quantum system $\mathcal{S}$ as required by quantum Darwinism cannot demonstrate that they are jointly observing $\mathcal{S}$ without a joint \textit{a priori} assumption of a classical boundary separating $\mathcal{S}$ from its environment $\mathcal{E}$.  Quantum Darwinism cannot, therefore, be regarded as providing a purely quantum-mechanical explanation of the ``emergence'' of classicality. 
\end{abstract}


\section{Introduction}

Quantum Darwinism (QD) \cite{zurek03rev, zurek06, zurek09rev} is an ambitious and detailed program to explain the ``emergence'' of an objective, classical world from minimal quantum mechanics alone.  The QD program has three components: (1) einselection of a pointer basis for any system of interest $\mathcal{S}$ by decohering interactions with the surrounding environment $\mathcal{E}$ \cite{zurek03rev, zurek93rev, zurek98rough}; (2) ``witnessing'' of the eigenstates of $\mathcal{S}$ in the einselected pointer basis by states of $\mathcal{E}$ \cite{zurek04, zurek05}; and (3) redundant encoding of the environmentally-witnessed eigenstates of $\mathcal{S}$ in multiple, disjoint fragments of $\mathcal{E}$ that are accessible to multiple independent observers \cite{zurek03rev, zurek06, zurek09rev, zurek04, zurek05, zurek07grand}.  Consistent with a conception of minimal quantum mechanics as an ``ultimate theory that needs no modifications to account for the emergence of the classical'' (\cite{zurek07grand}, p. 1), einselection, environmental witnessing and redundant encoding of pointer states are proposed by QD to fully account for the existence of ``objective'' physical properties, defined by Ollivier, Poulin and Zurek (OPZ) as follows:

\begin{quotation}
``A property of a physical system is \textit{objective} when it is:
\begin{list}{\leftmargin=2em}
\item
1. simultaneously accessible to many observers,
\item
2. who are able to find out what it is without prior knowledge about the system of interest, and 
\item
3. who can arrive at a consensus about it without prior agreement.''
\end{list}
\begin{flushright}
(p. 1 of \cite{zurek04}; p. 3 of \cite{zurek05})
\end{flushright}
\end{quotation}

That this OPZ definition of ``objective'' properties is intended to characterize the observable properties of not only microscopic quantum systems but also the macroscopic quantum systems composing the everyday world is stated explicitly: ``quantum Darwinism provides a satisfying explanation of the emergence of the objective classical world we perceive from the underlying quantum substrate'' (p. 19 of \cite{zurek05}); ``quantum Darwinism accounts for the transition from quantum fragility (of information) to the effectively classical robustness'' (p. 189 of \cite{zurek09rev}); ``[Q]uantum Darwinism ... explains the emergence of objectivity, as it allows many initially ignorant observers to independently obtain nearly complete information and reach consensus about the state of the system'' (p. 1 of \cite{zwolak09}; see also \cite{schloss04, schloss07, schloss08} and from a more philosophical perspective \cite{zurek98rough, zurek07grand, schloss08phil}).

This paper accepts the OPZ definition of objectivity as providing an operational definition of ``emergence'' as this term is commonly understood in discussions of the emergence of classicality from an underlying quantum ``reality'' described by fully-deterministic unitary dynamics (for review see \cite{schloss07}).  It then asks whether the physical mechanisms proposed by QD, namely einselection, witnessing of einselected eigenstates by the environment, and redundant encoding of environmentally-witnessed eigenstates are sufficient to render the ``properties'' of a macroscopic quantum system represented by its einselected eigenstates ``objective'' in the sense defined by OPZ.  This question is framed operationally: is it possible for two observers $\mathcal{O_{\mathit{1}}}$ and $\mathcal{O_{\mathit{2}}}$ to ``arrive at a consensus'' about the eigenstates of a macroscopic quantum system $\mathcal{S}$ given \textit{only} quantum-mechanical descriptions of $\mathcal{S}$ and its states and \textit{only} quantum-mechanical methods for discovering further information about $\mathcal{S}$, and is it possible for them to do this ``without prior knowledge'' and ``without prior agreement'' about $\mathcal{S}$ or its states?   It is shown that the answer to this operational question is ``no'': even if provided with a Hilbert-space decomposition $\mathcal{H_{S}} \otimes \mathcal{H_{E}}$, two observers $\mathcal{O_{\mathit{1}}}$ and $\mathcal{O_{\mathit{2}}}$ must by prior agreement share an assumption of sufficient information to specify the classical boundary separating $\mathcal{S}$ from its environment $\mathcal{E}$ in order to operationally \textit{identify} $\mathcal{S}$, and hence in order to arrive at a consensus that observable properties are properties of $\mathcal{S}$.  Such a classical boundary must be assumed even if the observers are allowed to directly manipulate and hence ``prepare'' $\mathcal{S}$ before making their observations.  A prior agreement specifying the classical boundary separating $\mathcal{S}$ from $\mathcal{E}$ amounts, however, to a shared assumption that $\mathcal{S}$ has already ``emerged'' as an objective system.  From this it is concluded that QD cannot, as formulated, account for the emergence of the classical world.

\section{Definitions and operational setting}

The OPZ definition of objectivity assumes the existence of a ``physical system'' $\mathcal{S}$ and at least two ``observers'' $\mathcal{O_{\mathit{1}}}$ and $\mathcal{O_{\mathit{2}}}$.  The meanings of the terms ``system'' and ``observer'' are left implicit in QD.  Zurek remarks at the end of his ``rough guide'' to decoherence that ``a compelling explanation of what the systems are - how to define them given, say, the overall Hamiltonian in some suitably large Hilbert space - would undoubtedly be most useful'' (p. 1818 of \cite{zurek98rough}).  In later papers introducing QD, Zurek assumes as ``axiom(o)'' of quantum mechanics that ``systems exist'' (p. 746 of \cite{zurek03rev}; p. 3 of \cite{zurek07grand}).  This emphasis on the existence of systems, together with the stated goal of explaining the emergence of an ``objective classical world'' quoted above, renders two variant interpretations of the OPZ definition highly implausible in the context of QD.  One variant interpretation is to regard ``objectivity'' as associated with ``properties'' \textit{only} and to regard ``systems'' as remaining non-objective.  
This variant interpretation of OPZ may be consistent with an instrumentalist interpretation of quantum mechanics (cf. \cite{bub04, fuchs10}), but in such an approach ``the real world ... is taken for granted'' (p. 7 of \cite{fuchs10}), so ``emergence'' is not a phenomenon to be explained and the OPZ definition itself is irrelevant.  A second variant interpretation is to regard the ``system'' as the universe as a whole, and hence as an already-objective entity for which the notion of ``emergence'' does not apply.  In this variant interpretation, however, there is no environment, so both the concept of decoherence and most of the formalism of QD are unneccessary.  In the present discussion, therefore, the OPZ definition is interpreted in the most straightforward and literal way, as providing an operational criterion for the ``emergence'' of \textit{both} properties and the systems that they characterize as ``objective''.  Interpreting the OPZ definition in this way and regarding QD as a proposed explanation of the ``emergence'' of both properties and systems into ``objectivity'' is consistent both with ``axiom(o)'' and with Zurek's statement that ``preferred states of $\mathcal{S}$ become objective'' by the mechanisms of QD (p. 182 of \cite{zurek09rev}).  For the present purposes, moreover, the systems of interest will be assumed to be \textit{macroscopic}.

``Observers'' are implicitly characterized by the OPZ definition as entities capable of arriving at a consensus, making prior agreements, having prior knowledge, and having ``access'' to the system $\mathcal{S}$.  Within QD, observers are distinguished from apparatus by their ability to ``readily consult the content of their memory'' (p. 759 of \cite{zurek03rev}) and are assumed to be much smaller than the environment $\mathcal{E}$ \cite{zurek03rev, zurek05, zurek09rev} but are otherwise uncharacterized.  The ``emergence'' of observers is not discussed explicitly; however, Zurek raises the question of whether QD is ``in some way behind the familiar natural selection'' (p. 188 of \cite{zurek09rev}), a question that has been pursued by Durt \cite{durt10}.  It is assumed here that observers are macroscopic physical systems and hence within the scope of the OPZ definition; under this assumption, circularity can be avoided provided that any pair of observers can conclude that any third observer is ``objective'' using the methods of QD within the restrictions on prior knowledge and prior agreement imposed by the OPZ definition.

Provided that the variant interpretation of $\mathcal{S}$ as the universe as a whole is ruled out as discussed above, the existence of the environment $\mathcal{E}$ is implicit in the OPZ definition.  In QD, $\mathcal{E}$ is assumed to be arbitrarily large (i.e. to have arbitrarily many quantum degrees of freedom) and to interact with any systems ``immersed'' in it, including $\mathcal{S}$, $\mathcal{O_{\mathit{1}}}$ and $\mathcal{O_{\mathit{2}}}$ \cite{zurek03rev, zurek05, zurek09rev}.  For the purposes of practical calculations, $\mathcal{E}$ is typically taken to comprise only specific kinds of degrees of freedom, e.g. the degrees of freedom of the ambient photon field (e.g. \cite{zurek10}).

The OPZ definition explicitly forbids ``prior knowledge'' and ``prior agreement'' on the part of $\mathcal{O_{\mathit{1}}}$ and $\mathcal{O_{\mathit{2}}}$.  However, certain prior agreements between $\mathcal{O_{\mathit{1}}}$ and $\mathcal{O_{\mathit{2}}}$ must be assumed in order to define the operational setting to which the OPZ definition refers.  In particular, $\mathcal{O_{\mathit{1}}}$ and $\mathcal{O_{\mathit{2}}}$ must recognize each other as observers in order to ``arrive at a consensus''; they must also be assumed to agree that they have been provided, e.g. by an unspecified oracle, with a single (quantum-mechanical) description of $\mathcal{S}$ or set of (quantum-mechanical) tools for identifying $\mathcal{S}$.  What is critical to the OPZ definition and to the project of QD as an explanation of the emergence of classicality from minimal quantum mechanics alone is that $\mathcal{O_{\mathit{1}}}$ and $\mathcal{O_{\mathit{2}}}$ have no prior shared \textit{classical} criteria with which to identify $\mathcal{S}$, since the availability of any shared classical criterion sufficient to identify $\mathcal{S}$ would imply that $\mathcal{S}$ had already ``emerged'' into classical objectivity.  Hence $\mathcal{O_{\mathit{1}}}$ and $\mathcal{O_{\mathit{2}}}$ cannot, for example, point to $\mathcal{S}$ and say ``\textit{that} is the system of interest.''  Indeed the project of QD is to explain \textit{how it is possible} for observers to point to anything and agree that it is ``objective'', i.e. that it has ``emerged'' from the quantum into the classical realm.

\section{Observation within the framework of QD}

The QD framework for describing the interaction between the observers $\mathcal{O_{\mathit{1}}}$ and $\mathcal{O_{\mathit{2}}}$ and the system $\mathcal{S}$ involves three distinct mechanisms.  First, $\mathcal{S}$ interacts with the environment $\mathcal{E}$ via a Hamiltonian $H_{\mathcal{S-E}}$.  This $\mathcal{S-E}$ interaction einselects as ``pointer'' states of $\mathcal{S}$ the eigenstates $\lbrace |s_{k}\rangle \rbrace$ of  $H_{\mathcal{S-E}}$ \cite{zurek03rev, zurek93rev, zurek98rough}.  Second, these eigenstates of $\mathcal{S}$ are ``witnessed'' and hence encoded in $\mathcal{E}$ \cite{zurek04, zurek05}.  Third, the encoding of $|s_{k}\rangle$ in $\mathcal{E}$ propagates outward from the locus of interaction into multiple disjoint fragments $\mathcal{F_{\mathit{i}}}$ of $\mathcal{E}$ under the action of the environmental self-Hamiltonian $H_{\mathcal{E}}$ \cite{zurek06, zurek09rev, zurek05}.  Each observer $\mathcal{O_{\mathit{i}}}$ is restricted to a single fragment $\mathcal{F_{\mathit{i}}}$.  It is assumed both that the $\mathcal{F_{\mathit{i}}}$ are mutually dynamically decoupled, and that information does not flow backwards from the $\mathcal{F_{\mathit{i}}}$ to $\mathcal{S}$.  The measurements carried out by $\mathcal{O_{\mathit{i}}}$ within $\mathcal{F_{\mathit{i}}}$ therefore have no effect either on any $\mathcal{F_{\mathit{j}}}$ with $j \neq i$ or on $\mathcal{S}$; hence both the outward propagation of the environmental encoding and the measurement interactions themselves can be considered classical \cite{zurek06, zurek09rev, zurek05}.  This assumption restricts observers to non-destructive measurements, and justifies use of a reduced density matrix formalism in which each $\mathcal{O_{\mathit{i}}}$ traces out all $\mathcal{F_{\mathit{j}}}$ with $j \neq i$.   Under these conditions, the degrees of freedom of $\mathcal{F_{\mathit{i}}}$ with which $\mathcal{O_{\mathit{i}}}$ interacts can be considered as an $\mathcal{O_{\mathit{i}}}$-specific macroscopic ``apparatus'' $\mathcal{A_{\mathit{i}}}$ that encodes the pointer state $|s_{k}\rangle$ of $\mathcal{S}$ by classical correlation.  As this ``apparatus'' is restricted by the einselection and environmental witnessing mechanisms to encoding only eigenstates of $H_{\mathcal{S-E}}$, it can be viewed as a classical physical implementation of an $\mathcal{O_{\mathit{i}}}$-specific generalized observable for $\mathcal{S}$ that incorporates the interaction $H_{\mathcal{S-E}}$.

In the context of the OPZ definition, this framework introduces what amounts to a prior agreement among the observers: the $\mathcal{O_{\mathit{i}}}$ agree to regard 
$\mathcal{E}$ as separable into the $\mathcal{F_{\mathit{i}}}$ plus the region immediately surrounding $\mathcal{S}$.  This is effectively an agreement that $\mathcal{E}$ enforces strong decoherence; any (quantum) information encoded only in some $\mathcal{F_{\mathit{j}}}$ with $j \neq i$ is effectively ``lost'' to $\mathcal{O_{\mathit{i}}}$.  Because this is a prior agreement about $\mathcal{E}$, not about $\mathcal{S}$, it is not ruled out by the OPZ definition's restrictions on prior agreements.  It is noteworthy, however, that this prior agreement renders the ``apparatus'' $\mathcal{A_{\mathit{i}}}$ with which $\mathcal{O_{\mathit{i}}}$ determines the state of $\mathcal{S}$ inaccessible to the other observers and hence \textit{nonobjective} under the OPZ definition.  This nonobjectivity of the observer-specific ``apparatus'' is the analog within QD of the ``personal'' probabilities for outcomes stressed by the quantum Bayesian approach \cite{fuchs10}.  It entails that observers can share outcomes but not observations.

It was shown in \cite{fields10} that under the above conditions of observation, and without further assumptions concerning $\mathcal{S}$, $\mathcal{O_{\mathit{1}}}$ and $\mathcal{O_{\mathit{2}}}$ cannot determine on the basis of non-destructive measurements and quantum-mechanical calculations alone whether indistinguishable outcomes $A_{1}$ and $A_{2}$ obtained from $\mathcal{A_{\mathit{1}}}$ and $\mathcal{A_{\mathit{2}}}$ respectively are redundant encodings of a single pointer state $|s_{k}\rangle$ of a single system $\mathcal{S}$, or encodings of distinct pointer states $|s_{k}\rangle$ and $|s^{\prime}_{k}\rangle$ of distinct systems $\mathcal{S}$ and $\mathcal{S}^{\prime}$ that merely happen, due to the choices of $\mathcal{A_{\mathit{1}}}$ and $\mathcal{A_{\mathit{2}}}$, to be indistinguishable.  Two observers $\mathcal{O_{\mathit{1}}}$ and $\mathcal{O_{\mathit{2}}}$ cannot, therefore, employ observations of indistinguishable and hence apparently redundant encodings in their accessible fragments of $\mathcal{E}$ to identify $\mathcal{S}$.  The reason for this is clear: restricted as they are to the $\mathcal{F_{\mathit{i}}}$, none of which contain degrees of freedom that interact directly with $\mathcal{S}$, the observers cannot get ``close enough'' to $\mathcal{S}$ to unambiguously demonstrate that the properties they are observing are properties of $\mathcal{S}$.  They are, therefore, unable to demonstrate that the environmental encodings of these properties are redundant; they must jointly \textit{assume} encoding redundancy in order to jointly identify $\mathcal{S}$, in violation of the OPZ restriction on ``prior agreement'' among observers.  Providing $\mathcal{O_{\mathit{1}}}$ and $\mathcal{O_{\mathit{2}}}$ with observer-specific implementations of a generalized observable for states of $\mathcal{S}$ is, without this assumption of encoding redundancy, insufficient to start the process of consensus-building contemplated by the OPZ definition of objectivity.

\section{Assuming encoding redundancy is equivalent to \\ assuming a classical boundary}

Given the need for a prior assumption of encoding redundancy, the operational question posed in the Introduction becomes the question of \textit{how much} information about $\mathcal{S}$ this prior assumption entails, and whether this information can reasonably be regarded as a ``given'' in a purely quantum mechanical theory that is intended to explain the emergence of a classical world.  In order to address this question, we consider what information $\mathcal{O_{\mathit{1}}}$ and $\mathcal{O_{\mathit{2}}}$ would require in order to demonstrate encoding redundancy.  The $\mathcal{A_{\mathit{i}}}$ with which the observers interact are by assumption \textit{classical} encodings, so demonstrating that the $\mathcal{A_{\mathit{i}}}$ are redundant encodings requires demonstrating that they are classically correlated.  The observers have no access to each other's $\mathcal{A_{\mathit{i}}}$, however, so they cannot use jointly-observed manipulations of the $\mathcal{A_{\mathit{i}}}$ to demonstrate their classical correlation.  They can demonstrate correlation between the $\mathcal{A_{\mathit{i}}}$ only by demonstrating, by calculation, that the $\mathcal{A_{\mathit{i}}}$ specifically, and hence jointly, encode the pointer states of $\mathcal{S}$.  Such a calculation requires the ability to calculate the pointer states of $\mathcal{S}$.  It is precisely such a calculation of pointer states that QD provides the formal tools to perform.  The question of interest is what must be assumed in order to use these tools.

Suppose that $\mathcal{O_{\mathit{1}}}$ and $\mathcal{O_{\mathit{2}}}$ are provided, as prior knowledge, with a Hilbert-space decomposition $\mathcal{H_{S}} \otimes \mathcal{H_{E}}$ that explicitly enumerates the degrees of freedom of $\mathcal{S}$ and implicitly specifies the degrees of freedom of $\mathcal{E}$.  For example, suppose that $\mathcal{S}$ is ``given'' as one mole of Au atoms, with position and excitation degrees of freedom, and $\mathcal{E}$ is specified to be the ambient visible-spectrum photon field.  From this specification, $\mathcal{O_{\mathit{1}}}$ and $\mathcal{O_{\mathit{2}}}$ can infer that the interaction $H_{\mathcal{S-E}}$ is visible photon scattering from Au atoms, can infer that the self-interaction $H_{\mathcal{S}}$ is at low enough energy to preserve atomic structure, and can assure themselves that $H_{\mathcal{S-E}}$ and $H_{\mathcal{S}}$ at least approximately commute as required by QD \cite{zurek05}.  However, this specification does not permit calculation of a unique set of pointer states for $\mathcal{S}$; it does not distinguish photon scattering from a fully-dispersed gas of Au from photon scattering from a solid block of Au.  Hence knowledge of the Hilbert-space decomposition is insufficient to demonstrate encoding redundancy in the $\mathcal{A_{\mathit{i}}}$, and therefore insufficient for a consensus between $\mathcal{O_{\mathit{1}}}$ and $\mathcal{O_{\mathit{2}}}$ that $\mathcal{S}$ is an objective system.

One of four distinct kinds of information is required, in addition to the Hilbert-space decomposition, to enable calculation of a set of expected pointer states of $\mathcal{S}$:

\begin{enumerate}
\item The full initial state vector for $\mathcal{S}$ in some appropriate basis, or the set of all amplitudes $\alpha_{i}$ with $|\alpha_{i}|^{2} \geq \epsilon$ for some suitable $\epsilon$.
\item A complete set of matrix elements $\langle s_{i}|H_{\mathcal{S}}|\mathit{s_{j}}\rangle$, or the set of all matrix elements larger than some suitable $\epsilon$.
\item A complete set of matrix elements $\langle s_{i}|H_{\mathcal{S-E}}|\mathit{e_{j}}\rangle$, or the set of all matrix elements larger than some suitable $\epsilon$.
\item A minimal closed macroscopic spatial boundary $\mathcal{B_{S}}$ such that all degrees of freedom of $\mathcal{S}$ can be regarded as inside $\mathcal{B_{S}}$ and all degrees of freedom of $\mathcal{E}$ can be regarded as outside $\mathcal{B_{S}}$.
\end{enumerate}

The first three kinds of information are purely quantum-mechanical, and hence involve no prior assumptions about the objectivity of $\mathcal{S}$.  However, none of these kinds of information can be obtained by observers restricted to environmental fragments $\mathcal{F_{\mathit{i}}}$ containing no degrees of freedom that interact directly with $\mathcal{S}$, i.e. they cannot be obtained within the observational framework of QD except by \textit{a priori} assumption.  The fourth kind of information is classical; the existence of a minimal closed macroscopic spatial boundary $\mathcal{B_{S}}$ straightforwardly violates the uncertainty principle.  Nonetheless, it is precisely this fourth kind of information that is actually used in typical QD calculations.  The most realistic available model of decoherence and pointer-state encoding by the photon field, for example, treats macroscopic dust particles as fully characterized by mass, permittivity constant and center-of-mass position \cite{zurek10}.  While the mass of a dust particle could be calculated from a specification of its quantum degrees of freedom (e.g. its atomic composition), its electrical permittivity and center-of-mass position depend on macroscopic boundary conditions that could only be determined by classical measurements. 

The assumption by observers of a classical boundary for $\mathcal{S}$ clearly violates the restrictions on ``prior agreement'' made by the OPZ definition, as such an assumption amounts to an assumption of classical objectivity.  In the absence of an oracle that provides $\mathcal{O_{\mathit{1}}}$ and $\mathcal{O_{\mathit{2}}}$ with either state-vector amplitudes or interaction matrix elements, $\mathcal{O_{\mathit{1}}}$ and $\mathcal{O_{\mathit{2}}}$ must interact directly with $\mathcal{S}$ to obtain the information necessary to demonstrate encoding redundancy.  Such interactions require a relaxation of the QD framework for observation, and in particular disallow the assumption of separability that justified the use of reduced density matrices and rendered the $\mathcal{A_{\mathit{i}}}$ effectively classical.  The manipulations of $\mathcal{S}$ required to determine either state-vector amplitudes or interaction matrix elements will in general not commute with $H_{\mathcal{S-E}}$, and manipulations performed by $\mathcal{O_{\mathit{1}}}$ will in general not commute with manipulations performed by $\mathcal{O_{\mathit{2}}}$.  Hence even if the observational framework of QD is relaxed to permit direct system-observer interactions, $\mathcal{O_{\mathit{1}}}$ and $\mathcal{O_{\mathit{2}}}$ cannot demonstrate to each other either that they are manipulating the same system $\mathcal{S}$, or that the system that they are manipulating is the same system that they were previously observing from a distance.  Suppose, for example, that $\mathcal{O_{\mathit{1}}}$ leaves her distant environmental fragment $\mathcal{F_{\mathit{1}}}$ to perform a ``preparation'' operation $P_{\mathit{1}}$ on $\mathcal{S}$ at some time $t$, with $[P_{\mathit{1}},H_{\mathcal{S-E}}] \neq 0$, while $\mathcal{O_{\mathit{2}}}$ remains in $\mathcal{F_{\mathit{2}}}$ and interacts only with $\mathcal{A_{\mathit{2}}}$ as before.  From the perspective of $\mathcal{O_{\mathit{1}}}$, the state of $\mathcal{S}$ immediately after the operation with $P_{\mathit{1}}$ is some eigenstate of $P_{\mathit{1}}$.  From the perspective of $\mathcal{O_{\mathit{2}}}$, the state of $\mathcal{S}$ at $t + \Delta t$, where $\Delta t$ is the decoherence time, is an eigenstate $|s_{m}\rangle$ of $H_{\mathcal{S-E}}$ distinct from the eigenstate $|s_{k}\rangle$ observed before $\mathcal{O_{\mathit{1}}}$'s operation with $P_{\mathit{1}}$.  This new eigenstate $|s_{m}\rangle$ cannot be observed by $\mathcal{O_{\mathit{1}}}$, who is interacting directly with $\mathcal{S}$, not with a distant environmental fragment.  In this situation, $\mathcal{O_{\mathit{1}}}$ and $\mathcal{O_{\mathit{2}}}$ are not making observations in the same basis, and hence cannot directly compare the states that they are observing; they therefore have even less evidence that they are observing the same system $\mathcal{S}$ than they had when they were restricted to distant but comparable observations from $\mathcal{F_{\mathit{1}}}$ and $\mathcal{F_{\mathit{2}}}$.  Faced with this inability to unambiguously identify $\mathcal{S}$ by observations, $\mathcal{O_{\mathit{1}}}$ and $\mathcal{O_{\mathit{2}}}$ are forced, if they are to conclude anything at all, to jointly \textit{assume} that they are manipulating and observing the same system.  This assumption is, however, equivalent to the original assumption of encoding redundancy.  Hence even if the restrictions on interaction with $\mathcal{S}$ imposed by QD are fully relaxed to permit arbitrary ``preparations'' of $\mathcal{S}$, $\mathcal{O_{\mathit{1}}}$ and $\mathcal{O_{\mathit{2}}}$ cannot \textit{demonstrate} encoding redundancy unless they abandon experiments and assume either state-vector amplitudes or interaction matrix elements \textit{a priori}, as given by an oracle.  Such complete quantum information is, however, not a reasonable ``given''; it is the information experimental science is supposed to obtain, not the information with which it is allowed to start.

In actual practice, a joint assumption by observers that they are manipulating the same system is typically formulated as an assumption that any manipulations carried out within a particular bounded region of space are manipulations of the system of interest.  As noted above, any such assumption violates the uncertainty principle, is therefore a classical assumption, and would be disallowed by the OPZ definition as an unacceptable ``prior agreement'' about $\mathcal{S}$.  It is clear, however, that any assumption that independent manipulations are manipulations of the same system, however formulated, is operationally equivalent to the assumption of a classical system boundary, a boundary inside of which the system being manipulated is assumed to be.  A classically demarcated and hence in some sense ``visible'' boundary justifies an assumption of encoding redundancy, and hence justifies the conclusion that observed effects are observed effects on the same system.  In the specific case of manipulations performed on macroscopic measurement apparatus, such assumptions are, as Bohr \cite{bohr28} made clear, the foundation upon which the experimental analysis of quantum states is based.

\section{Conclusion}

This paper considered an operational question: is it possible for two observers $\mathcal{O_{\mathit{1}}}$ and $\mathcal{O_{\mathit{2}}}$ to ``arrive at a consensus'' about the eigenstates of a macroscopic quantum system $\mathcal{S}$ given \textit{only} quantum-mechanical descriptions of $\mathcal{S}$ and its states and \textit{only} quantum-mechanical methods for discovering further information about $\mathcal{S}$, and is it possible for them to do this ``without prior knowledge'' and ``without prior agreement'' about $\mathcal{S}$ or its states, as required by the OPZ definition of objectivity?  The answer to this question is ``no'': observers must assume a classical boundary separating $\mathcal{S}$ from its environment $\mathcal{E}$ to even begin the process of characterizing $\mathcal{S}$.  This is true whether $\mathcal{O_{\mathit{1}}}$ and $\mathcal{O_{\mathit{2}}}$ are restricted to making observations of $\mathcal{S}$ ``from a distance'' as they are in QD, or allowed to interact directly with $\mathcal{S}$, i.e. to prepare $\mathcal{S}$ in various ways.

If a classical boundary for $\mathcal{S}$ must be assumed from the outset, $\mathcal{S}$ must be regarded as ``objective'', and in particular as objectively within its classical boundary, from the outset.  Hence the OPZ definition, by referring to a ``physical system'' that is ``accessible to many observers'' is assuming what it is attempting to operationally define, that is, classical objectivity.  The claim that QD provides an explanation of how systems can satisfy the OPZ definition of objectivity and hence ``emerge'' as classical is, therefore, too strong.  What QD in fact explains is how a system that has already emerged into classicality remains classical.  The demonstration by Zurek and Paz that decoherence enforces classicality on the chaotic orbit of Hyperion \cite{schloss08, zurek-paz94, zurek-paz96, zurek98hyp} exemplifies such a \textit{post hoc} explanation; a classical initial state for Hyperion with a classical spatial boundary at which decoherence acts is assumed, not calculated from quantum-mechanical first principles.  For macroscopic systems embedded in the environment, first-principles calculations to demonstrate an ``emergence'' into classicality cannot be made, not because an appropriate formalism is unavailable but because the information needed to employ this formalism cannot be obtained in an observational framework that does not permit classical assumptions.  Absent an ability to define them from quantum-mechanical first principles, macroscopic systems can be picked out as ``systems'' only by \textit{a priori} assumption, by taking them as ``given'' either by an oracle or by classical observations.  In his lifelong insistence on this point, Bohr \cite{bohr28} was right.

\section*{Acknowledgments}

The comments of anonymous referees on previous versions have contributed significantly to this paper.


\begin{thebibliography}{99}

\bibitem{zurek03rev} W. H. Zurek, Rev. Mod. Phys. 75 715 (2003). arXiv:quant-ph/0105.127v3

\bibitem{zurek06} R. Blume-Kohout and W. H. Zurek, Phys. Rev. A 73 062310 (2006). arXiv:quant-ph/0505031v2

\bibitem{zurek09rev} W. H. Zurek, Nature Phys. 5 181 (2009). arXiv:quant-ph/0903.5082v1

\bibitem{zurek93rev} W. H. Zurek, Prog. Theor. Phys. 89 281 (1993).

\bibitem{zurek98rough} W. H. Zurek, Phil. Trans. Royal Soc. London A 356 1793 (1998).

\bibitem{zurek04} H. Ollivier, D. Poulin and W. H. Zurek, Phys. Rev. Lett. 93 220401 (2004). arXiv:quant-ph/0307229v2

\bibitem{zurek05} H. Ollivier, D. Poulin and W. H. Zurek, Phys. Rev. A 72 042113 (2005). arXiv:quant-ph/0408125v3

\bibitem{zurek07grand} W. H. Zurek, arXiv:quant-ph/0707.2832v1 (2007).

\bibitem{zwolak09} M. Zwolak, H. T. Quan and W. H. Zurek, Phys. Rev. Lett. 103 110402 (2009). arXiv:quant-ph/0904.0418v2

\bibitem{schloss04} M. Schlosshauer, Rev. Mod. Phys. 76 1267 (2004). arXiv:quant-ph/0312059v4

\bibitem{schloss07} M. Schlosshauer, Decoherence and the Quantum to Classical Transition. (Springer, Berlin, 2007).

\bibitem{schloss08} M. Schlosshauer, Found. Phys. 38 796 (2008). arXiv:quant-ph/0605.249

\bibitem{schloss08phil} M. Schlosshauer and K. Camilleri, arXiv:quant-ph/0804.1609v1 (2008).

\bibitem{bub04} J. Bub, Stud. Hist. Phil. Mod. Phys. 35 241 (2004).

\bibitem{fuchs10} C. Fuchs, arXiv:quant-ph/1003.5209v1 (2010).

\bibitem{durt10} T. Durt, Found. Sci. 15 177 (2010).

\bibitem{zurek10} C. J. Reidel and W. H. Zurek, Phys. Rev. Lett. 105 020404 (2010). arXiv:quant-ph/1001.3419

\bibitem{fields10} C. Fields, Int. J. Theor. Phys. 49 2523 (2010). arXiv:quant-ph/1003.5136v2

\bibitem{zurek-paz94} W. H. Zurek and J. P. Paz, Phys. Rev. Lett. 72 2508 (1994).

\bibitem{zurek-paz96} W. H. Zurek and J. P. Paz, arXiv:quant-ph/9612.037v1 (1996).

\bibitem{zurek98hyp} W. H. Zurek, Phys. Scr. 1998 186 (1998). arXiv:quant-ph/9802.054v1

\bibitem{bohr28} N. Bohr, Nature 121 580 (1928).

\end{thebibliography}
\end{document}